\documentstyle[12pt]{article}
\begin{document}
\pagestyle{plain}
\begin{center}
DUAL AND DUAL-CROSS SYNCHRONIZATIONS IN CHAOTIC SYSTEMS.\\
E.M. Shahverdiev \footnote{Permanent address: Institute of Physics, 370143 Baku,Azerbaijan}
and K. A. Shore \footnote{Electronic address: alan@sees.bangor.ac.uk}\\
School of Informatics, University of Wales, Bangor, Dean Street, Bangor, LL57 1UT, Wales, UK\\
\end{center} 
~\\
In this paper we investigate dual synchronization and dual-cross synchronization between the transmitter and receiver, each one consisting of two master and slave systems. We demonstrate our approach using the example of  synchronization of chaotic external cavity laser diodes. In the synchronization scheme under consideration the 
joint error signal, which is the difference between the sum of the outputs from the master systems in the transmitter and the sum of outputs from the slave systems in the receiver, is fed into each slave system and synchronizations between the master lasers and slave lasers under possible configurations are studied. By use of error dynamics we derive existence conditions for synchronization between the transmitter and receiver. We find that in all the studied cases synchronization between the transmitter and receiver occurs, when the sum of the contributions from individual master (or slave) systems in the transmitter (or receiver) is equal to unity. Arbitrarily chosen master system from the transmitter synchronizes with those slave system from the receiver whose contribution weight to the joint error signal is equal to its contribution. These findings are considered to be of interest in multichannel communication schemes.
~\\
PACS number(s):05.45.-a, 05.45.Xt, 42.65.Sf, 42.55.Ah\\

~\\
PACS number(s):05.45.Xt, 42.65.Sf, 42.55.Ah\\
~\\
\indent The seminal papers[1] on chaos synchronization have stimulated a wide range of  
research activity: a recent comprehensive review of such work is found 
in [2]. Application of chaos control theory can be found in secure communications, optimization of nonlinear system performance and modeling 
brain activity and pattern recognition phenomena [2]. A particular 
focus of the work is on the development of secure optical communications 
systems based on control and synchronization of laser hyperchaos [3], because 
communication schemes on hyperchaos are considered to be more reliable. For 
practical use of this approach  particular emphasis is given to the use 
of chaotic external cavity semiconductor lasers, see e.g. [3], because laser systems 
with optical feedback are prominent representatives of time-delay 
systems, which can generate hyperchaos.\\
\indent Investigation of multichannel communication schemes, due to their 
practicality is of considerable research interest,see e.g. [4] and references therein.
Recent work treat the problem of simultaneous synchronization of two different 
pairs of chaotic oscillators with a single scalar signal, which it refers to as 
dual synchronization [5]: the outputs of a pair of master oscillators are linearly 
coupled and fed to a pair of slave oscillators.The signals from the slave 
oscillators are coupled in a similar way and subtracted from the signal received from 
the masters'. The difference signal (or joint error signal) is then injected into each slave oscillator. When the slaves are synchronized to their respective masters, the difference (error) signal is zero. Notice that there is no direct coupling between the two master oscillators and therefore the dual synchronization case is different from the problem of using a scalar signal to synchronize hyperchaotic or multidimensional oscillators.  In [5] the stability condition of dual synchronization is derived 
as $\epsilon_{1}+\epsilon_{2}=1$ (where $\epsilon_{1}$, $\epsilon_{2}$ are coupling parameters:$0\le \epsilon_{1}, \epsilon_{2}\le 1$) for a number of different classes of maps and for oscillators modelled by delay-differential equations by evaluating the Lyapunov exponent of the receiver with respect to the synchronized state.\\
\indent In this paper by use of error dynamics we derive the existence conditions of both dual and dual-cross synchronizations between the transmitter and receiver each one consisting of two 
master and slave systems. We demonstrate our approach using the example of synchronization of chaotic external cavity laser diodes. In dual synchronization as in [5] we synchronize the slave lasers to their respective master lasers. In 
dual-cross synchronization one synchronizes the slave lasers to their 
non-respective master lasers. We obtain that $\epsilon_{1}+\epsilon_{2}=1$ is also a existence condition for dual synchronization.
We also establish a existence condition for dual- cross synchronization: $\epsilon_{1}=\epsilon_{2}=\frac{1}{2}$.  We find that in all the cases studied here  synchronization between the transmitter and receiver occurs, when the sum of contributions from individual master (or slave) systems in the transmitter (or receiver) is equal to unity. Arbitrarily chosen master system from the transmitter synchronizes with those slave system from the receiver whose contribution weight to the joint error signal is equal to its contribution.\\
External cavity laser diodes are commonly modeled with the Lang-Kobayashi equations 
[6]. Consider the following systems of the Lang-Kobayashi 
equations for the real electric field amplitude E(t), slowly 
varying phase $\Phi(t)$ and the carrier number $n(t)$ for: the master 
laser pair $M1, M2$ 
$$\hspace*{2cm}\frac{dE_{M1}}{dt}=\frac{1}{2}G_{N} n_{M1}E_{M1} + k_{M1}E_{M1}(t-\tau )\cos (\omega_{0}\tau + \Phi_{M1}(t)-\Phi_{M1}(t-\tau))=F(t),\hspace*{0.5cm}$$
$$\hspace*{1.5cm}\frac{d\Phi_{M1}}{dt}=\frac{1}{2}\alpha G_{N}n_{M1}-k_{M1}\frac{E_{M1}(t-\tau )}{E_{M1}(t)}\sin (\omega_{0}\tau + \Phi_{M1}(t)-\Phi_{M1}(t-\tau))=F_{1}(t),\hspace*{0.5cm}$$
$$\hspace*{2cm}\frac{dn_{M1}}{dt}= (p-1)J_{th}-\gamma n_{M1}(t)-(\Gamma + G_{N}n_{M1})E_{M1}^{2},\hspace*{6.5cm}(1)$$

$$\hspace*{2cm}\frac{dE_{M2}}{dt}=\frac{1}{2}G_{N} n_{M2}E_{M2} + k_{M2}E_{M2}(t-\tau )\cos (\omega_{0}\tau + \Phi_{M2}(t)-\Phi_{M2}(t-\tau))=G(t),\hspace*{0.5cm}$$
$$\hspace*{1.5cm}\frac{d\Phi_{M2}}{dt}=\frac{1}{2}\alpha G_{N}n_{M2}-k_{M2}\frac{E_{M2}(t-\tau )}{E_{M2}(t)}\sin (\omega_{0}\tau + \Phi_{M2}(t)-\Phi_{M2}(t-\tau))=G_{1}(t),\hspace*{0.5cm}$$
$$\hspace*{2cm}\frac{dn_{M2}}{dt}= (p-1)J_{th}-\gamma n_{M2}(t)-(\Gamma + G_{N}n_{M2})E_{M2}^{2},\hspace*{6.5cm}(2)$$
and slave laser pair $S1, S2$
$$\hspace*{2cm}\frac{dE_{S1}}{dt}=\frac{1}{2}G_{N} n_{S1}E_{S1} + k_{S1}E_{S1}(t-\tau )\cos (\omega_{0}\tau + \Phi_{S1}(t)-\Phi_{S1}(t-\tau)) + e(t)=f(t) + e(t),$$
$$\hspace*{2cm}\frac{d\Phi_{S1}}{dt}=\frac{1}{2}\alpha G_{N}n_{S1}-k_{S1}\frac{E_{S1}(t-\tau )}{E_{S1}(t)}\sin (\omega_{0}\tau + \Phi_{S1}(t)-\Phi_{S1}(t-\tau))+e_{1}(t)=f_{1}(t)+e_{1}(t),$$
$$\hspace*{2cm}\frac{dn_{S1}}{dt}= (p-1)J_{th}-\gamma n_{S1}(t)-(\Gamma + G_{N}n_{S1})E_{S1}^{2},\hspace*{6.8cm}(3)$$
~\\
$$\hspace*{2cm}\frac{dE_{S2}}{dt}=\frac{1}{2}G_{N} n_{S2}E_{S2} + k_{S2}E_{S2}(t-\tau )\cos (\omega_{0}\tau + \Phi_{S2}(t)-\Phi_{S2}(t-\tau)) + e(t)=g(t) + e(t),$$
$$\hspace*{2cm}\frac{d\Phi_{S2}}{dt}=\frac{1}{2}\alpha G_{N}n_{S2}-k_{S2}\frac{E_{S2}(t-\tau )}{E_{S2}(t)}\sin (\omega_{0}\tau + \Phi_{S2}(t)-\Phi_{S2}(t-\tau))=g_{1}(t)+e_{1}(t),\hspace*{1cm}$$
$$\hspace*{2cm}\frac{dn_{S2}}{dt}= (p-1)J_{th}-\gamma n_{S2}(t)-(\Gamma + G_{N}n_{S2})E_{S2}^{2},\hspace*{6.8cm}(4)$$
here $G_{N}$ is the differential optical gain;$\tau$ is the  lasers' external 
cavity round-trip time;$\alpha$-the linewidth enhancement factor;
$\gamma$- the carrier decay rate;$\Gamma$-the cavity decay rate;$p$-
the pump current relative to the threshold value $J_{th}$ of the 
solitary laser;$\omega_{0}$ is the angular frequency of the solitary 
laser;$k$ is the feedback rate.\\
We define the joint error signal $e(t)$ to be fed into each slave laser as the difference signal between the sums of the master and 
slave lasers' outputs $u(t)$ and $v(t)$, respectively:
$$\hspace*{-3.5cm}e(t)=u(t)-v(t)=\epsilon_{1} F(t) + \epsilon_{2} G(t) -(\epsilon_{1}f(t)  + \epsilon_{2}g(t))$$
$$\hspace*{2cm}e_{1}(t)=u_{1}(t)-v_{1}(t)=\epsilon_{1} F_{1}(t) + \epsilon_{2} G_{1}(t) -(\epsilon_{1}f_{1}(t)  + \epsilon_{2}g_{1}(t)),\hspace*{4.5cm}(5)$$ 
Where, $\epsilon_{1}$ and $\epsilon_{2}$ are coupling parameters or contribution 
weights from the master lasers (or slave lasers) to the joint error signal $e(t)$.
In the following derivation  
of the existence conditions for dual and dual-cross synchronizations between chaotic laser pairs we will follow the approach developed in [7,8] using the investigation of the 
error dynamics. Earlier in [8] we have used this approach to derive a necessary (existence) condition for lag synchronization between unidirectionally coupled chaotic external cavity semiconductor lasers.\\
First consider the case of dual synchronization.
We write complete dual synchronization manifold as 
$$\hspace*{3cm}E_{M1}(t)=E_{S1}(t), n_{M1}(t)=n_{S1}(t), \Phi_{M1}(t)=\Phi_{S1} (t) (\bmod \hspace*{0.4cm} 2\pi),\hspace*{2.3cm}(6a)$$
$$\hspace*{3cm}E_{M2}(t)=E_{S2}(t), n_{M2}(t)=n_{S2}(t),\Phi_{M2}(t)=\Phi_{S2} (t) (\bmod \hspace*{0.4cm} 2\pi) .\hspace*{2.5cm}(6b)$$
In general, sychronization between master (driving) and 
slave (response) systems strongly depends on the driving variable. 
For example, as shown in the seminal paper by Pecora-Carroll for 
the paradigm Lorenz model driving by the $x$ and $y$ variables results  
in synchronization between master and slave systems, as the so-called 
conditional Lyapunov exponents for the response system are negative [1];
driving with the $z$ variable gives rise to only conditional 
synchronization, as one of the conditional Lyapunov exponents in this case 
is equal to zero, see, e.g. references [9].\\
Our similar analysis here shows that electric field amplitude is a good driving variable: indeed as can be seen from 
the third equations in eqs. (1)-(3) and (2)-(4), when the differences 
 $E_{M1}-E_{S1}$ and $E_{M2}-E_{S2}$ approach zero so do the differences 
between the corresponding carrier densities.\\ 
Now consider the deviation from the synchronization manifold  $e_{2}=E_{M1}-E_{S1}$ and $e_{3}=E_{M2}-E_{S2}$. Then writing eqs. (1-5) in the concise form 
(which indicate the applicability of the approach developed here to 
a wide class of continous dynamical systems)
of $\frac{dE_{M_{1}}}{dt}=F$, $\frac{dE_{M_{2}}}{dt}=G$,
$\frac{dE_{S_{1}}}{dt}=f + u - v$, $\frac{dE_{S_{2}}}{dt}=g + u -v$, and calculating errors' dynamics for $e_{2}$ and $e_{3}$ we arive at the system of equations:
$\frac{de_{2}}{dt}=F-f-(u-v)$, $\frac{de_{3}}{dt}=G-g-(u-v)$. Our goal is to  
achieve synchronization: $e_{2}=0$ and $e_{3}=0$. 
Then with $u=\epsilon_{1} F + \epsilon_{2} G$, $v=\epsilon_{1} f + \epsilon_{2} g$ one finds that \\ 
$$\hspace*{-6.2cm}(1-\epsilon_{1})(F-f)=\epsilon_{2}(G-g),$$
$$\hspace*{3cm}(1-\epsilon_{2})(G-g)=\epsilon_{1}(F-f). \hspace*{8.9cm}(7)$$
Using the expressions for the $F, G, f$ and $g$  from eqs. (1-4) on the synchronization manifolds (6), eqs. (7) can be modified to  
$$\hspace*{-3.2cm}(1-\epsilon_{1})(k_{M1}-k_{S1})E_{M1}=\epsilon_{2}(k_{M2}-k_{S2})E_{M2},$$
$$\hspace*{3cm}(1-\epsilon_{2})(k_{M2}-k_{S2})E_{M2}=\epsilon_{1}(k_{M1}-k_{S1})E_{M1}. \hspace*{5.9cm}(8)$$
From eqs.(8) we obtain that  
$$\hspace*{3cm}[(1-\epsilon_{1})(1-\epsilon_{2})-\epsilon_{1}\epsilon_{2}](k_{M1}-k_{S1})E_{M1}=0. \hspace*{6.2cm}(9)$$
Thus for $k_{M1,M2}$ $\neq$ $k_{S1,S2}$ we establish 
the relationship between the coupling weights of the master or slave lasers outputs to 
achieve dual synchronization:
$$\hspace*{6cm}\epsilon_{1} + \epsilon_{2}=1. \hspace*{9cm}(10)$$
It is straightforward to derive condition (10) also through investigation of the phase dynamics. We also observe that the condition (10), as mentioned above is fairly general for different types of chaotic continous dynamical systems. 
Indeed as follows from eqs.(7) for system of equations 
$\frac{dx_{1}}{dt}=F$, $\frac{dx_{2}}{dt}=G$ and 
$\frac{dE_{y_{1}}}{dt}=f + u - v$, $\frac{dy_{2}}{dt}=g + u -v$
with $u$ and $v$ defined above for $F$ $\neq$ $f$ and  $G$ $\neq$ $g$  
on the synchronization manifold $x_{1}=y_{1}$, $x_{2}=y_{2}$ one can easily arrive at the dual synchronization condition (10).\\
Next we consider dual-cross synchronization, when one synchronizes the slave lasers to their non-respective master lasers: $E_{M_{1}}=E_{S_{2}}$ and 
$E_{M_{2}}=E_{S_{1}}$. Following the procedure above leading to eqs.(7) and to condition (10), we find the existence condition for dual-cross synchronization:
$$\hspace*{6cm}\epsilon_{1} = \epsilon_{2}=\frac{1}{2}. \hspace*{8.6cm}(11)$$
Notice that condition (10) also holds for the dual-cross synchronization case.\\
Thus we conclude that both dual and dual-cross synchronization is possible if 
coupling parameters (or contribution weights) to the joint error signal from the oscillators to be synchronized become equal. In other words symmetry in contributions to the coupling plays a crucial role in the synchronization phenomenon studied here.
This conclusion is also valid for larger number of oscillators -constituent parts of the transmitter and receiver. Indeed if we have 
three master ($x_{1}(t) , x_{2}(t), x_{3}(t)$) and three slave ($y_{1}(t), y_{2}(t), y_{3}(t)$) oscillators in the transmitter and receiver, respectively, it is 
possible to establish that triple synchronization's  
 $x_{1}(t)=y_{1}(t)$, $x_{2}(t)=y_{2}(t)$, $x_{3}(t)=y_{3}(t)$
existence condition is $\epsilon_{1} + \epsilon_{2} + \epsilon_{3}=1$.
Triple-cross synchronization, e.g. $x_{1}(t)=y_{3}(t), x_{2}(t)=y_{1}(t), x_{3}(t)=y_{2}(t)$
takes place if 
$\epsilon_{1} = \epsilon_{2} = \epsilon_{3}=\frac{1}{3}$. In the case of three and more oscillators we also obtain the case of mixed synchronization, when some of the oscillators are coupled in crossed configuration, 
others- in the uncrossed configuration. For example, in the case of three oscillators in both the transmitter and receiver
we can define the following  synchronization manifold 
as a mixed synchronization manifold: $x_{1}(t)=y_{2}(t)$, $x_{2}(t)=y_{1}(t)$, $x_{3}(t)=y_{3}(t)$.
Again by use of error dynamics approach one obtains that the mixed synchronization's existence condition is: $2\epsilon +  \epsilon_{3}=1$ with $\epsilon_{1} = \epsilon_{2} = \epsilon$. In other words, we find that for all the types of synchronization it exists if the contribution weights to the joint error signal from the oscillators to be synchronized were equal. Generalization to the case of $N$ master and $N$ slave oscillators are straightforward.\\
To summarize, we have investigated dual synchronization and dual-cross synchronization between the transmitter and receiver each one consisting of two master and slave systems. By use of error dynamics we have derived the existence conditions for 
dual and dual-cross synchronizations. We have established that in all the studied cases synchronization occurs, when the sum of contribution weights from individual master (or slave) systems in the transmitter (or receiver) is equal to unity and arbitrarily chosen master system from the transmitter synchronizes with those slave system from the receiver whose contribution weight to the joint error signal is equal to its contribution. These findings might be interesting in multichannel communication schemes, because message decoding is dependent on synchronization between the transmitter and receiver.\\
Acknowledgements-This work is supported by the UK Engineering and Physical Sciences 
Research Council grant GR/R22568/01.\\

\end{document}